\documentclass[a4paper,11pt]{article}
\usepackage{bm,amsmath,amssymb,graphicx,bm}
\newcommand{\nc}{\newcommand}
\nc{\rnc}{\renewcommand}
\nc{\nn}{\nonumber}
\nc{\der}{{\partial}}
\rnc{\Im}{{\rm{Im}\,}}
\rnc{\Re}{{\rm{Re}\,}}
\nc{\db}{\displaybreak[0]\\}
\nc{\bra}{\langle}
\nc{\ket}{\rangle}

\nc{\lam}{\lambda}
\nc{\g}{{\mathfrak{g}}}
\nc{\zb}{\bar{z}}
\nc{\hb}{\bar{h}}
\nc{\J}{\mathcal{J}}
\nc{\su}{\widehat{\mathfrak{su}}(2)_k}

\newtheorem{conjecture}{Conjecture}

\numberwithin{equation}{section}
\numberwithin{lemma}{section}
\numberwithin{proposition}{section}
\numberwithin{theorem}{section}
\numberwithin{corollary}{section}
\numberwithin{conjecture}{section}

\begin{document}%
%
\title{Multiple Schramm-Loewner evolutions for conformal field theories with
Lie algebra symmetries}
\author{Kazumitsu Sakai\thanks{E-mail: sakai@gokutan.c.u-tokyo.ac.jp}\\\\
\it Institute of Physics, University of Tokyo, \\
\it Komaba 3-8-1,
Meguro-ku, Tokyo 153-8902, Japan \\\\
\\}
\date{}
\maketitle
%
\begin{abstract}
We provide multiple Schramm-Loewner evolutions (SLEs) to describe the scaling 
limit of multiple interfaces in critical lattice models possessing  Lie algebra 
symmetries. The critical behavior of the  models is described by 
Wess-Zumino-Witten (WZW) models. Introducing a multiple Brownian motion 
on a Lie group as well as that on the real line, we construct the multiple 
SLE with additional Lie algebra symmetries.  The connection between 
the resultant SLE and the WZW model can be understood via SLE 
martingales satisfied by the correlation functions in the WZW model.
Due to interactions among SLE traces, these Brownian motions 
have drift terms which are determined by partition functions 
for the corresponding WZW model. As a concrete example, we apply
the formula to the $\su$-WZW model. Utilizing the fusion rules in
 the model, we conjecture that there exists a one-to-one 
correspondence between the partition functions and the topologically
inequivalent configurations of the SLE traces. Furthermore, solving
the Knizhnik-Zamolodchikov equation, we exactly compute the probabilities 
of occurrence for certain configurations (i.e. crossing probabilities) 
of traces for the triple SLE.
\end{abstract}
%
\section{Introduction}
Geometric aspects of critical phenomena are characterized by random fractals
such as conformally  invariant fluctuations of local order parameters. 
They have been extensively studied from various different points of view,
especially in two dimensions (2D) where the conformal invariance imposes strong
constraints on the structure of critical phenomena.  Among them, the Schramm-Loewner
evolutions (SLEs) \cite{S1}, which directly describe geometric aspects of 2D critical phenomena
through simple 1D Brownian motions, 
have  brought a renewed interest in the theory of random fractals (see 
\cite{RS,Wer,KN,C1,Law,BB1,Gruz,Smir} for reviews).

\begin{figure}[ttt]
\begin{center}
\includegraphics[width=0.85\textwidth]{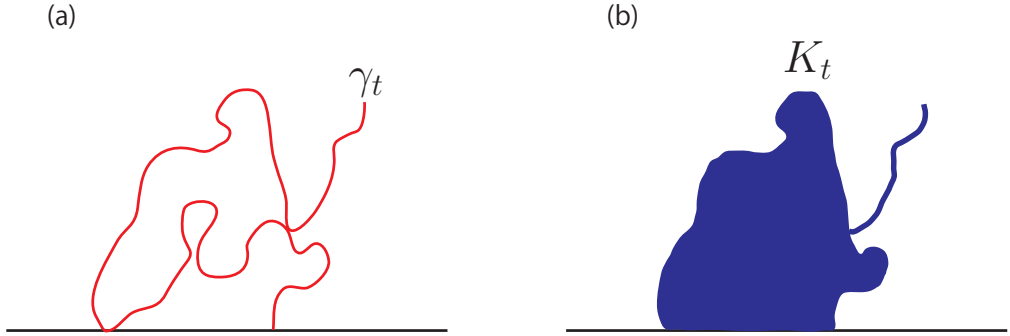}
\end{center}
\caption{Schematic view of a SLE trace ($\gamma_t$ denotes the tip of the trace) (a) 
and its hull $K_t$ (b).}
\label{hull}
\end{figure}

\begin{figure}[ttt]
\begin{center}
\includegraphics[width=0.85\textwidth]{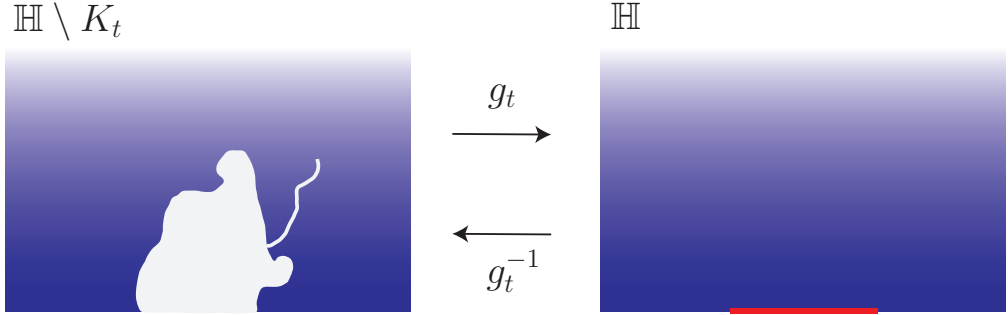}
\end{center}
\caption{Uniformizing map $g_t$: $\mathbb{H}\setminus K_t\to \mathbb{H}$ and 
its inverse. }
\label{uniform}
\end{figure}

The SLE is a stochastic process defined in the upper half plane $\mathbb{H}$.
Its evolution is described by the ordinary differential equation
\begin{equation}
d g_t(z)=\frac{2 dt}{g_t(z)-x_t}, \quad g_0(z)=z\in \mathbb{H},
\label{chordal}
\end{equation}
where  $x_t=\sqrt{\kappa} \xi_t$ is a Brownian motion on $\mathbb{R}$, starting at the origin
(i.e. $x_0=0$), and its expectation value and variance are given by $\mathbf{E}[d x_t]=0$ and 
$\mathbf{E}[dx_t dx_t]=\kappa dt$, respectively.  Here $\kappa>0$ is a diffusion coefficient
which essentially characterizes the SLE process.  The SLE \eqref{chordal} has a solution 
up to the explosion time $\tau_z$, i.e. the first time when $g_t(z)$ hits the singularity 
$x_t$. Let  $K_t=\overline{\{z\in \mathbb{H}| \tau_z<t \}}$ be the hull at time $t$
(see Fig.~\ref{hull} for a schematic view).
Then $K_t$ ($t\ge 0$) is an increasing family of hulls: $K_s \subset K_t$ for $s<t$.
Moreover $g_t(z)$ with $g_t(z)=z+O(1)$ at $z\to\infty$ (hydrodynamic normalization) 
is the unique conformal
map uniformizing the complement of the hull $K_t$ in the upper half plane $\mathbb{H}$:
$g_t(z): \mathbb{H}\setminus K_t \to \mathbb{H}$ (see Fig.~\ref{uniform}).
The image $x_t$ by $g_t^{-1}(x_t)$ 
defines the tip $\gamma_t$ of the growing random curve. More precisely, it is expressed
as $\gamma_t=\lim_{\epsilon\to +0} g_t^{-1}(x_t+i\epsilon)$.

%
The connection between the SLE \eqref{chordal} and conformal field theory (CFT) 
is well understood \cite{BB2,BB3,FW1,FW2,BB4,BB5,BB6,BBK,C1,BB1}.  
Specifically, it can be accomplished by noticing that
CFT correlation functions
\begin{equation}
\mathcal{M}_t=\frac{\bra \psi(\infty) \mathcal{O} \psi(\gamma_t)\ket}
{\bra \psi(\infty) \psi(\gamma_t) \ket}
\label{correlation}
\end{equation}
are SLE martingales: $d \mathcal{M}_t/dt=0$, where $\psi(\gamma_t)$ 
and $\psi(\infty)$ are boundary condition changing (bcc) operators with conformal
weights $h$,
which are inserted at the points $z=\gamma_t$ and $z=\infty$, respectively  
(see \cite{BB1,BBK} or next section for details). 
 Thus one can find the SLE corresponds
to the minimal conformal field theory $\mathcal{M}(p,p')$ ($p$, $p'$ are coprime 
integers satisfying the condition $p>p'\ge 2$) where the central charge and the conformal weights
of the primary fields are, respectively, given by \cite{BPZ,FMS}
\begin{equation}
c=1-6\frac{(p-p')^2}{p p'}, \quad h_{r,s}=\frac{(p r-p's)^2-(p-p')^2}{4p p'}.
\end{equation}
Then the diffusion constant $\kappa$ and the conformal weight $h$ of the boundary 
field $\psi$ are, respectively, expressed  as
\begin{equation}
c=\frac{(3\kappa-8)(6-\kappa)}{2\kappa}, \quad
h=\frac{6-\kappa}{2\kappa}=
\begin{cases}
h_{2,1} &\text{ for $\kappa=4p'/p\le 4$} \\
h_{1,2} &\text{ for $\kappa=4p/p'> 4$} 
\end{cases}.
\label{minimal}
\end{equation}
Namely the boundary field $\psi$ is degenerate at level two, i.e. possesses
a null field at level two.

An extension to the SLE connecting with  conformal field theories with Lie algebra
symmetries (i.e. WZW models \cite{N,W,KZ}) 
has been achieved by adding  the extra Brownian motion 
$\exp(\sum_a t^a d \theta_t^a(z))$ on a (semisimple) 
group manifold $G$ associated with a Lie algebra $\g$,  
where $t^a$'s ($a=1,\dots,\dim \g$) 
stand for any representation of the Lie algebra generators \cite{BGLW,ABI} 
(see   \cite{R1} for a very different approach).
The evolution of this additional stochastic process $\theta^a_t(z)$ is defined as
\begin{equation}
d\theta^a_t(z)=\frac{\sqrt{\tau}d \vartheta_t^a}{z-x_t}, \qquad
\mathbf{E}[d\vartheta_t^a]=0,\,\,  \mathbf{E}[d\vartheta_t^a d\vartheta_t^b]=\delta^{ab} dt 
 \,\,\,(\vartheta_t^a\in \mathbb{R}).
\label{lie}
\end{equation}
The combination of \eqref{chordal} and \eqref{lie} defines  a fractal curve 
living on the Lie group $G$ manifold. The SLE martingale $d \mathcal{M}_t/dt=0$, which 
should be satisfied by the CFT correlation function \eqref{correlation} 
(note that the bcc operators and the operator $\mathcal{O}$ take their 
values on $G$), determines the relation of the SLE defined by \eqref{chordal} and
\eqref{lie} with the corresponding 
WZW model. Some extensions to the SLE with other additional symmetries
have also been done in \cite{R2,R3,NR,Sant,Naza1,Naza2}.

In this paper we generalize the SLE for the system containing multiple 
random interfaces which possess additional Lie algebra symmetries, 
according to the theory developed in \cite{BBK}. Assuming that 
each SLE interface grows under an independent martingale in the infinitesimal 
time interval,  we extend \eqref{chordal} together with \eqref{lie} to the
case for the system with multiple interfaces. The evolution (cf. \eqref{chordal}
for the single case) characterizing the geometric aspect of the interfaces is described
by a multiple Brownian motion on the real line, while the evolution
(cf. \eqref{lie} for the single case) expressing the algebraic aspect is
described by  a multiple Brownian motion on the Lie group $G$. Both the Brownian
motions, however, have drift terms describing the interaction among the SLE
traces. Taking into account the SLE martingale, one finds that these drift terms
are determined by  the partition function in the corresponding WZW model. 
Moreover this partition function characterizes the configuration of the 
SLE traces. As an example, we apply our formula to the SLE for the 
$\su$-WZW models. Utilizing the fusion rules for $\su$, we conjecture
that there is a one-to-one correspondence between the partition 
functions and the topologically inequivalent configurations of the 
SLE traces. Further,  we exactly compute the probabilities of 
occurrence for certain configurations of traces (that correspond to 
crossing probabilities)  for  the triple SLE,
which can be obtained by solving the Knizhnik-Zamolodchikov (KZ) 
equation \cite{KZ}.

The paper is organized as follows. In the subsequent section, we  describe some
basic notions required in this paper. In section 3, the multiple SLE for conformal
field theory with Lie algebra symmetries is formulated. The drift terms of the 
driving Brownian motions are explicitly determined in section 4.  
A concrete application of the formula to the SLE for the $\su$-WZW model
is given in section 5.
\section{Preliminaries}
%
In this section, we describe several theoretical foundations required in 
subsequent sections.  In the former part of this section, we introduce SLE martingales
from the point of view of  statistical mechanics (see \cite{BB1} for details). 
SLE martingales are given by
CFT correlation functions involving bcc operators, and hence they become
a  key to decipher the relation between SLE and CFT.
In the latter part, general properties of correlation functions are 
explained in the case of the WZW model.
%
\subsection{SLE martingales and conformal correlators}
To formulate the SLEs correctly describing the behavior of 2D interfaces 
in critical systems, one must construct SLE martingales in terms of the 
corresponding statistical systems defined on $\mathbb{H}$.
Let $\bra \mathcal{O} \ket$ be the thermal average of an 
observable $\mathcal{O}$ defined in $\mathbb{H}$, and  
$\bra \mathcal{O} \ket|_{\{\gamma_t\}}$ be the thermal
average under a given shape of configuration of (multiple) 
interfaces, where $\{\gamma_t\}$ denotes the shape of configuration
with its occurrence probability $\mathbf{P}[{\{\gamma_t\}}]$.
Then the thermal average $\bra \mathcal{O} \ket$ must be given by 
\begin{equation}
\bra \mathcal{O}\ket=\mathbf{E}[\bra \mathcal{O} \ket|_{\{\gamma_t\}}]
=\sum_{\{\gamma_t\}} \mathbf{P}[\{\gamma_t\}]\bra \mathcal{O}\ket|_{\{\gamma_t\}},
\end{equation}
where the average $\mathbf{E}[\cdots]$ should be 
 taken over all the possible configurations.
The conditional expectation value 
$\bra \mathcal{O} \ket|_{\{\gamma_t\}}$ is thus time independent 
(i.e. conserved in mean), and
therefore it is a martingale. Here and in what follows, we denote it by
$\mathcal{M}_t$.

\begin{figure}[ttt]
\begin{center}
\includegraphics[width=0.85\textwidth]{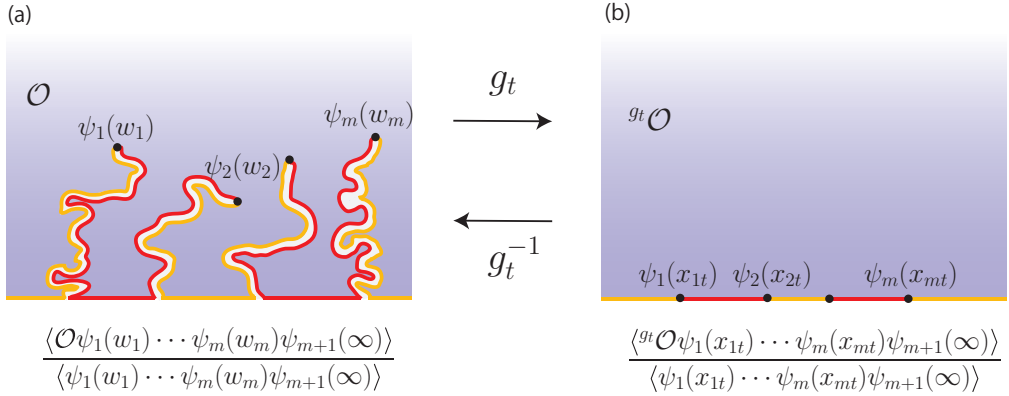}
\end{center}
\caption{CFT correlation functions on $\mathbb{H}\setminus\mathbb{K}$ (a). 
It can be transformed to the one defined on $\overline{\mathbb{H}}$
by the uniformizing map $g_t(z)$ (b).}
\label{corr}
\end{figure}

At the critical point where the conformal invariance is expected
in the system, the above observable can be described in terms of the CFT 
correlation functions. For the situation where the number of 
the interfaces under consideration is $m$, it reads
\begin{equation}
\mathcal{M}_t:=\bra \mathcal{O} \ket|_{\{\gamma_t\}}=
\frac{\bra \mathcal{O}\psi_1(w_1)\cdots\psi_m(w_m)\psi_{m+1}(\infty)\ket}
     {\bra \psi_1(w_1)\cdots\psi_m(w_m)\psi_{m+1}(\infty)\ket},
\label{correlation2}
\end{equation}
where $w_j$'s denote the positions of the tips of the interfaces.
Note that the CFT correlation functions are defined on the domain $\mathbb{H}$
removing the hull $K_t$, i.e. $\mathbb{H}\setminus K_t$ (see Fig.~\ref{corr}).
The operators $\psi_j(w_j)$'s inserted at the positions of the tips  denote 
primary bcc operators with  conformal weights $h$: under a local
conformal map $z\to z'= w(z)$, a primary field $\psi(z)$ transforms as 
$\psi(z)\to\psi'(z')$:
\begin{equation}
\psi'(z')=\left(\frac{dw(z)}{dz}\right)^{-h} \psi(z).
\label{CC}
\end{equation}
 The denominator in 
\eqref{correlation2} stands for the CFT partition function with a 
specific boundary condition fixed by the bcc operators $\psi_j(w_j)$.

Applying the conformal uniformizing map $g_t(z)$ 
(written with the same symbol as that used for the single SLE \eqref{chordal}), 
we  obtain 
\begin{equation}
\mathcal{M}_t=
\frac{\bra {}^{g_t}\mathcal{O}\psi_1(x_{1t})\cdots\psi_m(x_{mt})\psi_{m+1}(\infty)\ket}
     {\bra \psi_1(x_{1t})\cdots\psi_m(x_{mt})\psi_{m+1}(\infty)\ket}.
\end{equation}
Here $x_{jt}=g_t(w_j)$, and ${}^{g_t}\mathcal{O}$ is the
image of $\mathcal{O}$ by the map $g_t$. Note that the Jacobians coming
from the conformal map on $\psi_j$ have been canceled in
the numerators and the denominators\footnote{
The identity
$\bra \Phi'_1(z'_1)\cdots \Phi'_n(z'_n)  \ket=
\bra \Phi_1(z'_1)\cdots \Phi_n(z'_n) \ket$ holds for 
a global conformal map 
$z\to z'=f(z)$, where the field 
$\Phi_j(z_j)$ transforms as $\Phi_j(z_j)\to\Phi_j'(z_j')$
\cite{FMS}.  }.
Now SLE martingales $\mathcal{M}_t$ are expressed as the
CFT correlation functions on $\overline{\mathbb{H}}$, where
the bcc operators $\psi_j(x_j)$ are inserted at the points $x_j\in\mathbb{R}$
(see Fig.~\ref{corr}).

To proceed further, let us consider the case when the operator 
$\mathcal{O}$ is a product of an arbitrary number of 
primary fields $\mathcal{O}=\prod_{j=1}^n\phi_j(z_j,\zb_j)$ at positions 
$(z_j,\zb_j)$ and with conformal weights $(h_j,\hb_j)$.
By construction, the uniformizing map $g_t(z)$ can be analytically 
extended to the lower half plane: $g_t(\zb)=\bar{g_t}(z)$.
Then, the doubling trick can be applied to the CFT correlation functions.
The result reads
\begin{equation}
\mathcal{M}_t=
\prod_{j=1}^{2n}\left( \frac{\der y_{jt}}{\der z_j} \right)^{h_j}
\frac{\bra\prod_{j=1}^{2n} \phi_j(y_{jt}) \psi_1(x_{1t})\cdots\psi_m(x_{mt}) 
\psi_{m+1}(\infty)\ket}
{\bra\psi_1(x_{1t})\cdots\psi_m(x_{mt}) \psi_{m+1}(\infty)\ket }, 
\label{martingale}
\end{equation}
where we denote that $y_{jt}=g_t(z_j)$; $\phi_{j+n}=\bar{\phi}_{j}$;
$z_{j+n}=\zb_j=z_j^{\ast}$; $y_{j+n t}=\bar{y}_{jt}=y_{jt}^{\ast}$, 
$h_{j+n}=\hb_{j}$ for $1\le j \le n$.
Here $\phi_j(y_{jt})$ ($\bar{\phi}_{j}(y_{jt})$) stands for
the holomorphic (antiholomorphic) part  of the field 
$\phi_j(y_{jt},\bar{y}_{jt})$.

In this paper, we analyze the SLE martingales \eqref{martingale} for the
system that possesses additional Lie algebra symmetries. Namely we construct
the multiple SLE for the WZW models which are one of the most fundamental
CFTs with extra Lie algebra symmetries. In this case, the primary fields
constructing the SLE martingale \eqref{martingale} possess internal 
degrees of freedom, such as ``spin".

%
\subsection{WZW models and correlation functions}
%
Let us introduce several properties for correlation functions 
of WZW primary fields to analyze the SLE martingales \eqref{martingale}.
The WZW model is a CFT described by a field $g(z,\zb)$ taking values in 
a group manifold $G$ associated with a Lie algebra $\g$
\cite{FMS,N,W,KZ}.
The model is invariant under
\begin{equation}
g(z,\zb)\to\Omega(z)g(z,\zb)\bar{\Omega}^{-1}(\zb),
\label{GG}
\end{equation}
where $\Omega$ and $\bar{\Omega}(\zb)$ denote arbitrary
matrices valued in $G$.  This invariance gives rise to Noether currents
$J(z)=-k \der_z g g^{-1}$ and 
$\bar{J}(\zb)=k g^{-1}\der_{\zb} g$,
which can be written as
\begin{equation}
J(z)=\sum_{a=1}^{\dim\g} J^a(z) t^a,\quad
\bar{J}(\zb)=\sum_{a=1}^{\dim\g} \bar{J}^a(\zb) t^a,
\end{equation}
where $t^a$'s stand for any matrix representation of the generator
of $\g$, with commutation relations $[t^a,t^b]=\sum_c i f^{ab}_{\,\,\,\,\,\,c} t^c$.
The parameter $k$ is a positive integer referred to as the level.
Hereafter we only consider the holomorphic components, as if there 
were no boundary (cf. \eqref{martingale}) \cite{C2}.  Let $\bra X \ket$
be a correlator of $G$-valued fields. Then the infinitesimal transformation
$\Omega(z)=1+\omega(z)$ leads to the Ward identity
\begin{equation}
\delta_\omega \bra X \ket=-\frac{1}{2\pi i}\oint dz \sum_{a=1}^{\dim\g} \omega^a(z) 
\bra J^a(z) X\ket.
\label{ward-alg}
\end{equation}

On the other hand, conformal aspects of the WZW model are described by the stress energy 
tensor $T(z)$:
\begin{equation}
T(z)=\frac{1}{2(k+h^{\vee})}\sum_{a=1}^{\dim \g}
:J^a(z) J^a(z):,
\label{sugawara}
\end{equation}
where $h^{\vee}$ denotes the dual Coxeter number of $\g$. Note 
that $:\,\,:$ means the ``normal ordering" defined as
\begin{equation}
:A(z)B(z):=\frac{1}{2\pi i}\oint_{z}\frac{dw}{w-z} A(w)B(z).
\end{equation}
The infinitesimal conformal transformation $z\to z'=z+\epsilon(z)$ leads to the
Ward identity
\begin{equation}
\delta_\epsilon\bra X \ket=-\frac{1}{2\pi i}\oint dz \epsilon(z)\bra T(z) X\ket.
\label{ward-geo}
\end{equation}
This geometric part of the Ward identity \eqref{ward-geo} together with the
algebraic part \eqref{ward-alg} are a key ingredient in analyzing the SLE
martingale.

The current $J^a(z)$ and the stress energy tensor $T(z)$ can be expanded in terms
of the modes $J^a_n$ and $L_n$, respectively.  Namely
\begin{equation}
J^a(z)=\sum_{n\in \mathbb{Z}} J_n^a z^{-n-1}, \quad
T(z)=\sum_{n\in \mathbb{Z}} L_n z^{-n-2},
\label{mode}
\end{equation}
where $J_n^a$ and $L_n$ denote, respectively, the 
generators\footnote{Here we use the orthonormal basis in terms
of the Killing form: $K(J_n^a,J_m^b)=\delta_{n+m,0}\delta^{a,b}$.
In that case, the structure constants can be written as
$f^{ab}_{\,\,\,\,\,\,c}=f_{abc}$, where $f_{abc}$ is 
antisymmetric in all three indices.} of the 
affine Lie algebra $\hat{\g}$ with level $k$, and the generators
of the Virasoro algebra:
\begin{align}
&[J^a_n, J^b_m]=\sum_c i f^{ab}_{\,\,\,\,\,\,c} J^c_{n+m}+k n \delta^{a,b}\delta_{n+m,0},
                 \nn  \\
&[L_n,J_m^a]=-m J_{n+m}^a, \nn \\
& [L_n,L_m]=(n-m)L_{n+m}+\frac{c}{12}(n^3-n)\delta_{n+m,0}.
\label{Vir}
\end{align}
The central charge $c$ of the Virasoro algebra is then given by
\begin{equation}
c=\frac{k \dim \g}{k+h^{\vee}}.
\label{central}
\end{equation}
By construction \eqref{sugawara}, the Virasoro generators are not independent of 
the affine generators:
\begin{equation}
L_n=\frac{1}{2(k+h^{\vee})}\sum_{a=1}^{\dim\g}\sum_{m\in \mathbb{Z}}
 :J_m^a J_{n-m}^a:,
\label{L-J}
\end{equation}
where the normal ordering $:\,\,:$ means that the operator with  larger index $n$
is placed at the rightmost position.

The primary fields in the WZW model are defined as the fields transforming 
covariantly with
respect to the $G(z)$ transformation \eqref{GG}: $\delta_\omega g=\omega g$. 
Together with the conformal
covariance (see \eqref{CC}), these properties can be expressed in terms of the operator 
product expansions (OPE) via the Ward identities \eqref{ward-geo} and \eqref{ward-alg}:
\begin{equation}
T(z)\psi_{\lambda}(w)=\frac{h_\lambda \psi_\lambda(w)}{(z-w)^2}+
    \frac{\der_w \psi_\lambda(w)}{z-w}+\text{reg.}, \quad 
J^a(z) \psi_{\lambda}(w)=\frac{-t_\lambda^a \psi_\lambda(w)}{z-w}+\text{reg.},
\label{OPE}
\end{equation}
where the field $\psi_\lam(z)$ takes values in the representation specified by the 
highest weight $\lambda$, and $t_\lambda^a$ is the generator $t^a$ 
in that representation.
 Furthermore utilizing the field-state correspondence, i.e. 
$|\psi_\lambda\ket:=\lim_{z\to0}\psi_\lambda(z)|0\ket$, we can translate these 
properties into
\begin{alignat}{2}
& L_0 |\psi_\lambda\ket=h_\lambda |\psi_\lambda\ket, \quad && L_n|\psi_\lambda\ket=0 
\,\,\,(n>0),  \nn \\
&J_0^a |\psi_\lambda\ket=-t_\lambda^a |\psi_\lambda\ket, \quad && 
J_n^a |\psi_\lambda\ket=0 \,\,\, (n>0).
\label{FS}
\end{alignat}
Insertion of the relation \eqref{L-J} into the l.h.s. of the first equation in the above yields
\begin{equation}
L_0|\psi_\lambda\ket=\frac{1}{2(k+h^{\vee})}\sum_{a=1}^{\dim \g}
 J_0^a J_0^a|\psi_\lambda\ket=
\frac{1}{2(k+h^{\vee})}\sum_{a=1}^{\dim\g} t_\lam^a t_\lam^a|\psi_\lambda\ket
=\frac{(\lam,\lam+2\rho)}{2(k+h^{\vee})}|\psi_\lambda\ket.
\label{primary}
\end{equation}
In the last equality we used the explicit form of the eigenvalue of the 
quadratic Casimir. The quantity $\rho$ denotes the Weyl vector, i.e. 
the sum of the fundamental weights $\rho=\sum_{i=1}^r \Lambda_i$,
where $r$ is  the rank of $\g$. Comparing the r.h.s. in the
first equation in \eqref{FS} with the above,  we find 
\begin{equation}
h_\lam=\frac{(\lam,\lam+2\rho)}{2(k+h^{\vee})}.
\label{wt}
\end{equation}

Since the Virasoro generators are expressed in terms of 
affine generators as \eqref{L-J}, the other arbitrary states 
are of the form $J_{-l_1}^a J_{-l_2}^b \cdots |\psi_\lambda\ket$ with 
$l_1, l_2\dots$ positive integers.
Let $(L_{-l}\psi_\lambda)(z)$ and $(J_{-l}^a \psi_\lam)(z)$ be the descendent fields
corresponding to the states $L_{-l}|\psi_\lam\ket$ and $J_{-l}^a |\psi_\lam\ket$, respectively.
Combining the OPE \eqref{OPE} and the mode expansions \eqref{mode}, one
finds that the correlation functions $\bra (L_{-l}\psi_\lam)(z) X\ket$ and 
$\bra (J_{-l}^a \psi_\lam)(z) X\ket$, where $X=\prod_{i=1}^m \psi_{\lam_i}(z_i)$,
satisfy the following equations:
\begin{alignat}{2}
&\bra (L_{-l}\psi_\lam)(z) X\ket=\mathcal{L}_{-l} \bra \psi_\lam(z) X \ket,
\quad &&\mathcal{L}_{-l}=\sum_{j=1}^m \left[\frac{(l-1)h_{\lam_j}}{(z_j-z)^l}
-\frac{\der_{z_j}}{(z_j-z)^{l-1}}\right], \nn \\
&\bra (J_{-l}^a \psi_\lam)(z) X\ket=\mathcal{J}_{-l}^a\bra \psi_\lam(z) X \ket,
\quad && \mathcal{J}_{-l}^a=\sum_{j=1}^m \frac{t_{\lam_j}^a}{(z_j-z)^l}.
\label{LJ-op}
\end{alignat}
Note  that the global $G$-invariance requires 
$\delta_\omega \bra \psi_\lam(z) X \ket=0$. Using the
Ward identity \eqref{ward-alg} with constant $\omega$
and  the OPE \eqref{OPE}, we obtain a relation satisfied 
by the correlation function:
\begin{equation}
\sum_{a=1}^{\dim\g}\left(t_\lam^a+\sum_{j=1}^m t_{\lam_j}^a\right)
\bra \psi_\lam(z) X \ket=0.
\label{global}
\end{equation}
This constraint together with the global conformal invariance
(or equivalently $SL(2,\mathbb{C})$-invariance) 
fix the structure of the two- and three-points
correlation functions.

To close this section, 
let us derive a crucial equation called Knizhnik-Zamolodchikov (KZ) 
equation \cite{KZ} satisfied by the correlation functions of the WZW 
primary fields. The constraint stems from the fact
that the Virasoro generators are not independent of the affine 
generators as in \eqref{L-J}.  
For $n=-1$, we have $L_{-1}=\sum_{a=1}^{\dim\g} J_{-1}^a J_0^a/(k+h^{\vee})$. 
Then a null state $|\chi\ket=0$ is given by
\begin{equation}
|\chi\ket=\left(L_{-1}+\frac{1}{k+h^{\vee}}\sum_{a=1}^{\dim\g} t_\lam^a 
J_{-1}^a\right) |\psi_\lam\ket=0,
\end{equation}
where we used the property in \eqref{FS}. Using the field-state correspondence and 
inserting the property \eqref{LJ-op} into the correlation functions $\bra \chi(z) X\ket$,
one obtains the KZ equation:
\begin{equation}
\left(\der_{z}-\frac{1}{k+h^{\vee}}\sum_{a=1}^{\dim\g}
\sum_{j=1}^m\frac{t_\lam^a t_{\lam_j}^a}{z-z_j}\right)
\bra \chi(z) X\ket=0.
\label{kz}
\end{equation}
Here we used the translation invariance  
$(\der_z +\sum_{j=1}^m \der_{z_j}) \bra \chi(z) X\ket=0$,
which can be easily verified from \eqref{ward-geo} and the OPE \eqref{OPE} 
by setting  $\epsilon(z)=\epsilon$.
%
\section{Multiple SLEs for WZW models}
%
Now we generalize the SLE \eqref{chordal} and \eqref{lie} for the 
system containing multiple random interfaces with additional Lie algebra symmetries.
In the infinitesimal time interval, we expect that each SLE interface grows under an 
independent martingale. Let us discuss the case where the number of the interfaces is $m$.
Then the uniformizing map $g_t(z)$ with the hydrodynamic normalization may be of the form
\cite{BBK} (see also \cite{C3,Dub,Gra,KL} for other approaches).
\begin{equation}
dg_t(z)=\sum_{\alpha=1}^m \frac{2 dq_\alpha}{g_t(z)-x_{\alpha t}}, \quad g_0(z)=z,
\label{geo-SLE}
\end{equation}
where $dq_\alpha$'s mean  infinitesimal time intervals satisfying
the condition $\sum_{\alpha=1}^m dq_\alpha=dt$.
The random processes $x_\alpha$ ($1\le\alpha\le m$), which play a role as driving
forces for the growth of interfaces, should be written as
the It\^o stochastic differential equations:
\begin{equation}
dx_{\alpha t}=\sqrt{\kappa}d\xi_{\alpha t}+d F_{\alpha t},
\label{geo-driving}
\end{equation}
where $\xi_{\alpha t}$ is  an $\mathbb{R}^m$-valued  
Brownian motion whose expectation 
value and variance are, respectively, given by 
\begin{equation}
\mathbf{E}[d {\xi}_{\alpha t}]=0, \quad
\mathbf{E}[d\xi_{\alpha t} d\xi_{\beta t}]=\delta_{\alpha\beta}dq_\alpha.
\label{br1}
\end{equation}
Namely $dq_\alpha$ prescribes the growth rate of each interface.
The quantity $d F_{\alpha t}$ denotes a drift term proportional to $d q_\alpha$,
which comes from interactions among interfaces, and will be determined
later by the SLE martingale.

To extend \eqref{geo-SLE} to evolutions with Lie algebra symmetries,
we define a stochastic process $\exp[\sum_a d\theta_t^a(z) t^a]$ ($a=1,\dots,\dim\g$)
living on a Lie group 
manifold $G(z)$, where $\theta^a_t(z)$ is written as 
\begin{equation}
d\theta^a_t(z)=\sum_{\alpha=1}^m\frac{dp^a_{\alpha t}}{z-x_{\alpha t}}, 
\quad \theta^a_0(z)=0
\label{alg-SLE}
\end{equation}
with
\begin{equation}
dp_{\alpha t}^a=\sqrt{\tau} d\vartheta_{\alpha t}^a+dG_{\alpha t}^a .
\label{alg-driving}
\end{equation}
Here $\vartheta_{\alpha t}^a$ is  an 
$\mathbb{R}^{m \dim\g}$-valued  Brownian motion with
\begin{equation}
\mathbf{E}[d\vartheta^a_{\alpha t} ]=0, \quad
\mathbf{E}[d\vartheta^a_{\alpha t} d\vartheta_{\beta t}^b]=
\delta^{ab}\delta_{\alpha\beta}dq_\alpha,
\label{br2}
\end{equation}
and $dG_{\alpha t}^a$ stands for a drift term 
proportional to $dq_\alpha$, which will  also be determined later.
For $z=x_{\beta t}$, we must define
\begin{equation}
d\theta_t^a(x_{\beta t})=\sum_{\substack{\alpha=1\\ \alpha\ne\beta}}^m
\frac{dp^a_{\alpha t}}{x_{\beta t}-x_{\alpha t}}.
\end{equation}

The growth of interfaces with affine Lie algebra symmetries
is described by both the geometric  \eqref{geo-SLE} and the 
algebraic  \eqref{alg-SLE}  components.

\section{Drift terms and SLE martingales}
The remaining problem is to determine the drift terms appearing 
in the driving forces \eqref{geo-driving} and in the Brownian motion \eqref{alg-driving}.
To achieve it, we must evaluate the variation of the SLE martingale $\mathcal{M}_t$
involving WZW primary fields. For the WZW models, the SLE martingale is written as
\begin{equation}
\mathcal{M}_t=
\prod_{j=1}^{2n}\left( \frac{\der y_{jt}}{\der z_j} \right)^{h_{\mu_j}}
\frac{\bra\prod_{j=1}^{2n} \phi_{\mu_j}(y_{jt}) \psi_{\lam_1}(x_{1t})\cdots
\psi_{\lam_m}(x_{mt}) \psi_{\lam_{m+1}}(\infty)\ket}
{\bra\psi_{\lam_1}(x_{1t})\cdots\psi_{\lam_m}(x_{mt}) \psi_{\lam_{m+1}}(\infty)\ket },
\label{mart-SLE}
\end{equation}
where $y_{jt}=g_t(z_j)$ and $x_{\alpha t}=g_t(w_\alpha)$ 
(see section~2.1 for details). Note that the primary fields 
$\phi_{\mu_j}$ and $\psi_{\lam_\alpha}$
constructing the correlation function take values in the
representation specified by the highest weights $\mu_j$ and $\lam_\alpha$,
respectively.

The drift terms are determined
by the condition which makes $\mathcal{M}_t$ to be a martingale.
To simplify the notations  we sometimes omit the index $t$, $\lam$ and $\mu$
(e.g. $x_{\alpha t}=x_\alpha$, $\theta_t^a=\theta^a$, $\psi_{\lam_\alpha}(x_\alpha)=
\psi_\alpha(x_\alpha)$,
$h_{\lam_j}=h_j$,
etc.), if there is no ambiguity. 
Let $Z_t$, $Z_t^\phi$ and $J_t^\phi$ be the denominator,
numerator and Jacobian factor of \eqref{mart-SLE}, respectively.
First we explicitly compute $dJ_t^\phi$.  A simple manipulation
leads to
\begin{equation}
\der_{q_\alpha}\left( \frac{\der y}{\der z} \right)^{h}=
h \left( \frac{\der y}{\der z} \right)^{h-1}
\der z\left(\frac{\der y}{\der q_\alpha}\right)  
=
h\left( \frac{\der y}{\der z} \right)^{h}\der y
\left(\frac{\der y}{\der q_\alpha}\right) 
=-\left( \frac{\der y}{\der z} \right)^{h}
\frac{2h}{(y-x_\alpha)^2},
\label{jacob}
\end{equation}
where in the last equality, we applied the geometric component of the SLE
\eqref{geo-SLE}. Thus we obtain the variation of the Jacobian factor:
\begin{equation}
d J_t^\phi=-J_t^\phi\sum_{\alpha=1}^m
\sum_{j=1}^{2n}
\frac{2h_j}{(y_j-x_\alpha)^2} dq_\alpha.
\label{der-Jacobi}
\end{equation}

Next we shall calculate  $d (J_t^\phi Z_t^\phi)$.  
The It\^o derivative of the bulk fields $\phi(y)$
is given by
\begin{align}
d \phi(y)&=\frac{\der \phi(y)}{\der y}\sum_{\alpha=1}^m\frac{\der y}{\der q_\alpha} 
dq_\alpha+\sum_{a=1}^{\dim\g} t^a d\theta^a(y)
              \phi(y)
+\frac{1}{2}\sum_{a=1}^{\dim\g} t^a d\theta^a(y) 
                   \sum_{b=1}^{\dim\g} t^b d\theta^b(y)
   \phi(y)                      \nn \\
&=\sum_{\alpha=1}^m\left[ \frac{2dq_\alpha\der_y}{y-x_\alpha}+
     \sum_{a=1}^{\dim\g}
     \frac{dp_\alpha^a t^a}{y-x_\alpha}+
    \frac{\tau}{2}\sum_{a=1}^{\dim\g}
     \frac{dq_\alpha t^a t^a}{(y-x_\alpha)^2}
\right]\phi(y).
\end{align}
Here we applied \eqref{geo-SLE} to the first term, and
\eqref{alg-SLE} to the second and third terms.
For the third term we  also used  
the property \eqref{br2}.
Similarly, for the boundary fields $\psi_{\alpha}(x_\alpha)$, we obtain
\begin{align}
d\psi_{\alpha}(x_\alpha)&=
\left[dq_\alpha\frac{\kappa}{2}\der_{x_\alpha}^2+dx_\alpha\der_{x_\alpha}  
+\sum_{a=1}^{\dim\g} t^a d\theta^a(x_\alpha)
+\frac{1}{2}\sum_{a=1}^{\dim\g} t^a d\theta^a(x_\alpha) 
                   \sum_{b=1}^{\dim\g} t^b d\theta^b(x_\alpha)
\right]
   \phi_{\alpha}(x_\alpha)           
\nn \\
&=\left[dq_\alpha\frac{\kappa}{2}\der_{x_\alpha}^2+
d x_\alpha \der_{x_\alpha}
+ \sum_{a=1}^{\dim\g}\sum_{\substack{\beta=1\\ \beta\ne\alpha}}^m
 \left(\frac{dp_\beta^a t_{\alpha}^a}{x_\alpha-x_\beta}+
    \frac{\tau}{2}
     \frac{dq_\beta t_\alpha^at_\alpha^a}{(x_\alpha-x_\beta)^2}\right)
\right]\psi_{\alpha}(x_\alpha).
\end{align}
The above two relations together with \eqref{der-Jacobi} give the 
form of $d(J_t^\phi Z_t^\phi)$.
Explicitly it reads
\begin{align}
\frac{d(J_t^\phi Z_t^\phi)}{J_t^\phi}=&
\sum_{\alpha=1}^m dq_\alpha\left(
\frac{\kappa}{2}\mathcal{L}_{\alpha,-1}^2-2\mathcal{L}_{\alpha,-2}+
\frac{\tau}{2}\sum_{a=1}^{\dim\g}\J_{\alpha,-1}^a\J_{\alpha,-1}^a
\right)  Z_t^\phi\nn \\
&
+\sum_{\alpha=1}^m \left(
2\widetilde{\mathcal{L}}_{\alpha,-2} dq_\alpha
+\mathcal{L}_{\alpha,-1} dx_\alpha
+\sum_{a=1}^{\dim\g}\J_{\alpha,-1}^a dp_\alpha^a \right) Z_t^\phi,
\label{num}
\end{align}
where we define 
\begin{align}
&\mathcal{L}_{\alpha,-l}=\sum_{j=1}^n 
\left(
\frac{(l-1)h_j}{(y_j-x_\alpha)^l}-\frac{\der_{y_j}}{(y_j-x_\alpha)^{l-1}}
\right)+
\sum_{\substack{\beta=1\\\beta\ne\alpha}}^m
\left(
\frac{(l-1)h_\beta}{(x_\beta-x_\alpha)^l}-\frac{\der_{x_\beta}}{(x_\beta-x_\alpha)^{l-1}}
\right), \nn \\
&\J_{\alpha,-l}^{a}=\sum_{j=1}^n\frac{t_{\lam_j}^a}{(y_j-x_\alpha)^l}
+\sum_{\substack{\beta=1\\\beta\ne\alpha}}^m
\frac{t_{\lam_\beta}^a}{(x_\beta-x_\alpha)^l},
\label{LJ-op2}
\end{align}
and $\widetilde{\mathcal{L}}$  (resp. $\widetilde{\J}$) is given by subtracting
the first sum depending on $y_j$ from the r.h.s. of the first (resp. second) 
equation in \eqref{LJ-op2}. Note that the operator $\mathcal{L}$ (resp. $\J$) 
characteristically appears in the correlation functions between a
descendent field $(L_{-l} \psi_{\lam_\alpha})(x_\alpha)$ 
(resp. $(J_{-l} \psi_{\lam_\alpha})(x_\alpha)$) and some composite 
primary field (see \eqref{LJ-op}). To derive \eqref{num}, we have also
applied the identity $\der_{x_\alpha} Z_t^\phi=\mathcal{L}_{\alpha,-1}Z_t^\phi$,
which comes from the translation invariance of the correlation functions
(see the explanation below \eqref{kz}).

In  completely the same manner, $d Z_t$ can also be evaluated:
\begin{align}
dZ_t=&
\sum_{\alpha=1}^m dq_\alpha\left(
\frac{\kappa}{2}\widetilde{\mathcal{L}}_{\alpha,-1}^2-2\widetilde{\mathcal{L}}_{\alpha,-2}+
\frac{\tau}{2}\sum_{a=1}^{\dim\g}\widetilde{\J}_{\alpha,-1}^a
 \widetilde{\J}_{\alpha,-1}^a
\right)  Z_t\nn \\
&
+\sum_{\alpha=1}^m \left(
2\widetilde{\mathcal{L}}_{\alpha,-2} dq_\alpha
+\widetilde{\mathcal{L}}_{\alpha,-1} dx_\alpha
+\sum_{a=1}^{\dim\g}\widetilde{\J}_{\alpha,-1}^a dp_\alpha^a \right) Z_t,
\label{den}
\end{align}

By construction, we must set $h_{\lam_\alpha}=h$ ($1\le \alpha\le m$) for
the conformal weights of the bcc operators. Furthermore, due to the
constraint between the conformal weight $h_\lambda$ 
and the representation $\lambda$ \eqref{wt}, all the bcc operators
$\psi_{\lam_\alpha}(x_\alpha)$ except for $\psi_{\lam_{m+1}}(\infty)$
must take values in the representation specified by 
$\lam$, or its conjugate $\lam^{\ast}$ (note that 
$h_\lam=h_{\lam^{\ast}}$ holds). The conformal weight $h_{\lam_{m+1}}$ for the
bcc operator inserted at $\infty$ is determined by fusion rules.
Thanks to this restriction, we can determine, in principle,
the relation of the parameters $\kappa$, $\tau$ and the weight $h_\lam$ of the
bcc operators by considering the case that there is only a single interface ($m=1$)
as in \cite{BGLW,ABI}. We see this in the next subsection. 
\subsection{Single case ($m=1$)}
%
For the single SLE (set $\alpha=1$ and $m=1$),  one sees $d Z_t=0$ due to the 
translation invariance of the two-point correlation functions 
(it can also be obtained directly from \eqref{den} by setting $m=1$). 
Therefore the SLE martingale  gives a constraint to
the correlation function $Z_t^\phi$, i.e.  $\mathbf{E}[d(J_t^\phi Z_t^\phi)]=0$.
From the relation \eqref{num} and the definitions \eqref{geo-driving} and \eqref{alg-driving},
one easily see that this restriction leads to $dF_1=0$, $dG_1^a=0$ and
\begin{equation}
\left(
\frac{\kappa}{2}\mathcal{L}_{\alpha,-1}^2-2 \mathcal{L}_{\alpha,-2}
+\frac{\tau}{2}
\sum_{a=1}^{\dim\g}\J_{\alpha,-1}^{a}\J_{\alpha,-1}^{a}\right)Z_t^\phi=0 
\quad (\alpha=1).
\label{l2}
\end{equation}

Namely any drift terms do not show up in the driving forces \eqref{geo-driving} and
\eqref{alg-driving}, as expected. Moreover the constraint \eqref{l2} indicates that the bcc operators
must have  null states at level 2.  Namely $\mathcal{M}_t$ is a martingale if
and only if the bcc operators have the following null states at level 2:
\begin{equation}
0=|\chi\ket:=\left(\frac{\kappa}{2}L_{-1}^2-2 L_{-2}+\frac{\tau}{2} \sum_{a=1}^{\dim \g}
J_{-1}^a J_{-1}^a\right)|\psi_{\lam_1}\ket.
\label{l2-null}
\end{equation}
Here $L_n$ and $J_n^a$ are, respectively, 
the Virasoro and affine generators \eqref{Vir}.
This is equivalent to the conditions $J_1^b |\chi\ket=0$ and $J_2^b|\chi\ket=0$, 
which respectively lead to
\begin{equation}
\left[
(\tau k-\tau h^{\vee}-2)J_{-1}^b+\kappa J_0^b L_{-1}+i\tau 
\sum_{a,c}f^{ab}_{\,\,\,\,\,\,c}J_0^a J_{-1}^c\right]|\psi_{\lam_1}\ket=0,
\quad
(\kappa +\tau h^{\vee}-4)J_0^b|\psi_{\lam_1}\ket=0.
\label{ns-condition}
\end{equation}
This necessary and sufficient condition is rather involved. For some simple
cases (e.g. $\su$), 
however, we can directly solve the above equations by
acting the generator $J_1^d$. (Note that more elegant procedure utilizing
the KZ-equation has been developed in \cite{ABI}.) 

For later convenience, here
we explicitly write down the results for the $\su$-WZW
model. The  central charge $c$ 
\eqref{central} and the conformal weight 
$h_{\lam_1}$  \eqref{wt} of the bcc operator $\psi_{\lam_1}(x_1)$ in
the spin-$j/2$ representation are, respectively,  written as
\begin{equation}
c=\frac{3k}{k+2}, \qquad h_{\lam_1}=h_{j \Lambda}=\frac{j(j+2)}{4(k+2)},
\label{weight}
\end{equation}
where $\Lambda$ denotes the fundamental weight, and we used $h^{\vee}=2$.
{}From the direct evaluation of \eqref{ns-condition} for $\su$ case,
one finds that the conditions in \eqref{ns-condition} are valid 
only for the case that the bcc operator carries spin-1/2 ($j=1$) \cite{ABI}. 
Then \cite{BGLW,ABI}
\begin{equation}
\kappa=\frac{4(k+2)}{k+3}, \qquad 
\tau=\frac{2}{k+3} \quad (k\ge 2).
\label{kappa}
\end{equation}
For $k=1$, $\kappa$ and $\tau$ can not be specified, and only the relation
$\kappa+2\tau=4$ is imposed.
This case, however, corresponds to a $c=1$ CFT (cf. \eqref{minimal}), 
and therefore we shall set $\kappa=4$.
%
\subsection{Multiple case ($m>1$)}
%
Now  we identified
the bcc operators $\psi_{\lam_j}(z_j)$. 
Namely they have null states at level 2 and must satisfy the 
condition \eqref{l2-null}. Utilizing the field-state correspondence,
one finds that the first sums in \eqref{num} and \eqref{den}  vanish. 

Thus the drift term of the It\^o derivative of the CFT correlation function 
\begin{equation}
d\mathcal{M}_t=d\left(\frac{J_t^\phi Z_t^\phi}{Z_t}\right)
=\frac{d(J_t^\phi Z_t^\phi)}{Z_t}-\frac{J_t^\phi Z_t^\phi dZ_t}{Z_t^2}
-\frac{d(J_t^\phi Z_t^\phi )dZ_t}{Z_t^2}+
\frac{J_t^\phi Z_t^\phi (dZ_t)^2}{Z_t^3}
\end{equation}
is explicitly given by
\begin{align}
\mathbf{E}[d\mathcal{M}_t]=&J_t^\phi \sum_{\alpha=1}^m\left(dF_\alpha -
\kappa dq_\alpha  \der_{x_\alpha}\log Z_t
-2\sum_{\substack{\beta=1\\\beta\ne \alpha}}^m
\frac{dq_\beta }{x_\alpha-x_\beta}
\right)\der_{x_\alpha}\left(\frac{Z_t^\phi}{Z_t}\right)  \nn \\
&\qquad+
\frac{J_t^\phi}{Z_t}
\sum_{\alpha=1}^m\sum_{a=1}^{\dim\g}
\left[
\left(dG_\alpha^a-
\frac{\tau dq_\alpha}{Z_t}\widetilde{J}^{a}_{\alpha,-1}Z_t
\right)\left(J_{\alpha,-1}^a Z_t^\phi-
\frac{Z_t^\phi\widetilde{\J}_{\alpha,-1}^a}{Z_t}Z_t\right)
\right],
\end{align}
where we substituted the relations \eqref{geo-driving} and \eqref{alg-driving}.
By recalling that $\mathcal{M}_t$ is the SLE martingale, the above quantity must be
zero. Thus one finds the drift terms $d F_\alpha$ and $dG_{\alpha}^a$ are
described by the partition function $Z_t$:
\begin{align}
&dF_\alpha =\kappa dq_\alpha \der_{x_\alpha} \log Z_t+
2\sum_{\substack{\beta=1\\\beta\ne \alpha}}^m
\frac{dq_\beta }{x_\alpha-x_\beta}, \quad
dG_\alpha^a =\frac{\tau}{Z_t}\sum_{\substack{\beta=1\\\beta\ne\alpha}}^m
\frac{t_{\lam_\beta}^a Z_t}{x_\beta-x_\alpha}dq_\alpha.
\end{align}
\subsection{Main claim}
To summarize, we have constructed the multiple SLEs for $\hat{\mathfrak{g}}_k$-WZW 
models:
\begin{alignat}{2}
&dg_t(z)=\sum_{\alpha=1}^m \frac{2 dq_\alpha}{g_t(z)-x_{\alpha t}},&\qquad&
d x_{\alpha t}=\sqrt{\kappa}d\xi_{\alpha t}+dF_{\alpha t}  \nn \\
& d\theta_t^a(z)=\sum_{\alpha=1}^m\frac{dp^a_{\alpha t}}{z-x_{\alpha t}},
&& dp_{\alpha t}^a=\sqrt{\tau} d\vartheta_{\alpha t}^a+dG_{\alpha t}^a 
\quad (1\le a\le\dim(\g)).
\label{sle-sum}
\end{alignat}
The drift terms $d F_{\alpha t}$ and $dG_{\alpha t}^a$ are given by
\begin{align}
&dF_{\alpha t} =\kappa dq_\alpha \der_{x_{\alpha t}} \log Z_t+
2\sum_{\substack{\beta=1\\\beta\ne \alpha}}^m
\frac{dq_\beta }{x_{\alpha t}-x_{\beta t}}, \quad
dG_{\alpha t}^a =\frac{\tau}{Z_t}\sum_{\substack{\beta=1\\\beta\ne\alpha}}^m
\frac{t_{\lam_\beta}^a Z_t}{x_{\beta t}-x_{\alpha t}}dq_\alpha,
\label{driving-sum}
\end{align}
where $\xi_{\alpha t}$ (resp. $\vartheta_{\alpha t}^a$) is the
$\mathbb{R}^m$ (resp. $\mathbb{R}^{m \dim\g}$)-valued  Brownian motion 
whose expectation value and variance are given by 
\begin{alignat}{2}
&\mathbf{E}[d \xi_{\alpha t}]=0, \quad &&
\mathbf{E}[d\xi_{\alpha t} d\xi_{\beta t}]=\delta_{\alpha\beta}dq_\alpha, \nn \\
&\mathbf{E}[d\vartheta^a_{\alpha t} ]=0, \quad &&
\mathbf{E}[d\vartheta^a_{\alpha t} d\vartheta_{\beta t}^b]=
\delta^{ab}\delta_{\alpha\beta}dq_\alpha.
\end{alignat}
$Z_t$ is a partition function involving the bcc operators:
\begin{equation}
Z_t=
\bra\psi_{\lam_1}(x_{1t})\cdots\psi_{\lam_m}(x_{mt}) 
\psi_{\lam_{m+1}}(\infty)\ket,
\label{CFT-cor}
\end{equation}
where all the bcc operators
$\psi_{\lam_\alpha}(x_\alpha)$ except for $\psi_{\lam_{m+1}}(\infty)$
must take values in  representations specified by 
$\lam$, or its conjugate $\lam^{\ast}$. The conformal weight $h_{\lam_{m+1}}$ for the
bcc operator inserted at $\infty$ is determined by fusion rules (see next section
for example).
The structure of the
partition function is described by both the global $G$-invariance (cf. \eqref{global}),
and the KZ-equation (cf. \eqref{kz})
\begin{equation}
\sum_{a=1}^{\dim\g} \sum_{\alpha=1}^{m+1} t_{\lam_\alpha}^a Z_t=0,
\quad
\left(\der_{z_\alpha}-\frac{1}{k+h^{\vee}}
\sum_{a=1}^{\dim\g}
\sum_{\substack{\beta=1\\\beta\ne\alpha}}^{m}
\frac{t_{\lam_\alpha}^a t_{\lam_\beta}^a}{z_\alpha-z_\beta}\right)
Z_t=0.
\label{global2}
\end{equation}
%
\section{Multiple SLEs for $\su$-WZW models}
%
%
As a concrete application of our formula, let us consider 
multiple SLEs for the $\su$-WZW model where the bcc operators
carry spin-1/2, and discuss  topologies for the SLE traces.

For the $\su$ case, the fusion rules of the primary fields
are similar to those in the minimal CFTs. Therefore it is
natural to extend the argument \cite{BBK} (see also \cite{Gra,KL}) 
describing the geometric configurations of the SLE traces in minimal CFTs 
to the $\widehat{\mathfrak{su}}(2)_k$-WZW model.
Thus we make the following conjecture.

\begin{conjecture} \label{conj-config}
There exists a 
one-to-one correspondence between topologically inequivalent 
configurations of $\su$ multiple SLE traces and the independent 
solutions of the KZ-equation satisfied by the partition 
functions.
\end{conjecture}

\begin{figure}[ttt]
\begin{center}
\includegraphics[width=0.7\textwidth]{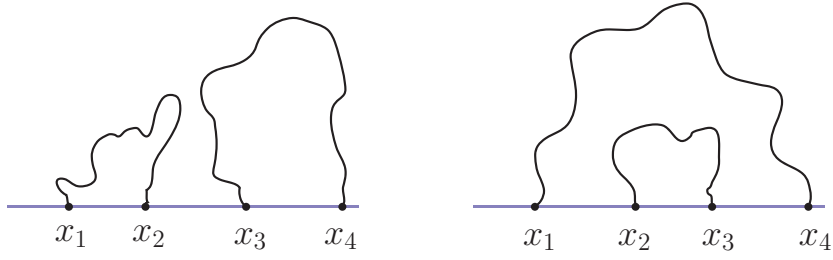}
\end{center}
\caption{Configurations  for $m=4$ and $n=2$. 
There are two ($c_{4,2}=\binom{4}{2}-\binom{4}{1}=2$) topologically 
inequivalent configurations.}
\label{config}
\end{figure}

We describe this conjecture more specifically. 
Let $x_\alpha$ ($1\le\alpha\le m$)  be the positions 
where the $m$ SLE traces start to grow, and be  ordered as 
$x_1<\cdots<x_m<\infty$. Consider the case that the 
traces eventually form  $m-n$ disjoint curves in $\mathbb{H}$ 
so that each point $x_\alpha$ is an end point of exactly one curve and  
$\infty$ is an end point of exactly $m-2n\ge 0$ curves.  Namely 
$n$ disjoint curves form arches (more precisely $n$ pairs of growing 
curves hit each other's tips and consequently form $n$ arches)
and other $m-2n$ curves converge toward 
the point at $\infty$ (see Fig.~\ref{config} for $m=4$ and $n=2$).
Then the number of topologically inequivalent configurations are given
by
\begin{equation}
c_{m,n}=\binom{m}{n}-\binom{m}{n-1}.
\end{equation} 
This is nothing but a Kostka number appearing as 
the coefficient of the irreducible decomposition for the 
$m$ tensor product of the $\mathfrak{su}(2)$ 
fundamental representation $L_{\Lambda}^{\otimes m}$ 
into $L_{(m-2n)\Lambda}$ ($L_{j \Lambda}$ denotes 
the integral representation with the highest weight 
$j \Lambda$, where $j\in \mathbb{Z}_{\ge 0}$ and
$\Lambda$ is the fundamental weight of $\mathfrak{su}(2)$): 
\begin{equation}
L_\Lambda^{\otimes m}=\bigoplus_{n=0}^{\lfloor m/2 \rfloor} 
c_{m,n}L_{(m-2n)\Lambda},
\label{su2}
\end{equation}
where $\lfloor x \rfloor$ denotes the integer part of $x$.

Now we mention that the relation between the 
geometric configurations and the CFT partition functions $Z$ \eqref{CFT-cor} 
(hereafter we omit the index $t$ to simplify the notation). To this end, 
we consider the fusions of the 
$\su$ primary fields:
\begin{equation}
\psi_{j_1 \Lambda} \otimes \psi_{j_2 \Lambda}=
\bigoplus_{\substack{j_3=|j_1-j_2|\\ 
       j_1+j_2+j_3\equiv 0 \mod 2}}^{\min(j_1+j_2,2k-j_1-j_2)}
       \psi_{j_3 \Lambda},
\label{fusion}
\end{equation}
where $\psi_{j \Lambda}$ stands for the $\su$ bcc primary field 
taking  values in the integral representation $L_{j \Lambda}$.
For sufficiently large $k$, the above rules reduce to
the decomposition \eqref{su2}.
Applying the fusion procedures \eqref{fusion} to the bcc operators
$\psi_{\lam_\alpha}$ ($1\le\alpha\le m$) recursively, one can reduce
$Z$ to a two point function involving the  fusion operator and
the bcc operator at $\infty$. Thus, for the non-vanishing partition
functions, the conformal weight $h_{\lam_{m+1}}$ of the bcc
operator $\psi_{\lam_{m+1}}(\infty)$ must be equivalent to that 
of the fusion operator. Fig.~\ref{fusion-fig} shows the 
possible conformal weight $h_{\lam_{m+1}}$ up to $m=4$. Due to the
fusion procedures \eqref{fusion},  for sufficiently large $k$, 
one finds that the number  of the paths from the left to 
$h_{(m-2n)\Lambda}$ corresponds to the number of topologically 
distinct configurations for the SLE traces, i.e. $c_{m,n}$.
 For generic  $k$, the number
of paths is, in general, constrained by the fusion rules,
which affects the structure of the partition functions,
and therefore the realization of the geometric configurations.
%

For instance, the configuration where all the curves eventually 
converge toward 
$\infty$ (i.e. no arches) corresponds to the path to the weight 
$h_{\lam_{m+1}}=h_{m \Lambda}$. By a standard argument for the  
CFT correlation functions, one obtains the corresponding 
partition function $Z$:
\begin{equation}
Z=
 \prod_{j<k}^m (x_j-x_k)^{\frac{1}{2(k+2)}} \quad \text{ for $m \le k$}.
\end{equation}
One can easily check that this partition function satisfies the KZ-equation \eqref{global2}.
The  factorized correlation function with the same exponents $1/(2(k+2))$
does not exist for $m> k$,
indicating that the no-arch configuration is allowed only 
when $m \le k$.

%
\subsection{Double SLEs}

\begin{figure}[t]
\begin{center}
\includegraphics[width=0.4\textwidth]{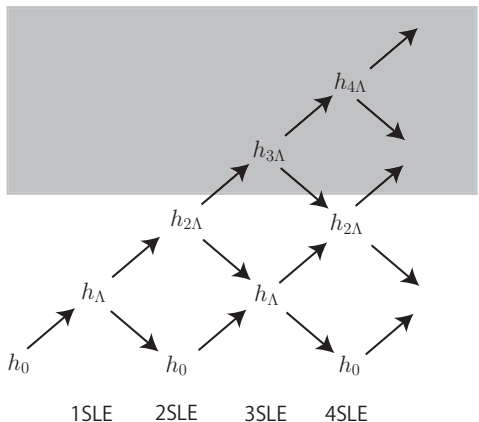}
\end{center}
\caption{The fusion rules of the bcc operator $\psi_{\Lambda}$. 
The primary fields with weight $h_{\lambda \Lambda}$ for $\lambda>k$
do not exist due to the fusion rules. For instance, for $k=2$,
the primary fields in the shadow area vanish.}
\label{fusion-fig}
\end{figure}

According to the argument  in \cite{BBK}, 
we confirm our conjecture by considering the
simplest case ($m=2$), where only two SLE traces exist
in the upper half plane $\mathbb{H}$. To analyze specifically,
let us write down the partition function $Z$. Up to a constant
factor, it is given by
\begin{equation}
Z=\bra \psi_{\lam_3}(\infty) \psi_\Lambda(x_1) \psi_{\Lambda}(x_2)\ket
=\lim_{x\to\infty} x^{2h_{\lam_3}}
\bra \psi_{\lam_3}(x) \psi_\Lambda(x_1) \psi_{\Lambda}(x_2)\ket=
(x_1-x_2)^\Delta,
\end{equation}
where the exponent is $\Delta=h_{\lam_3}-2h_\Lambda$. 
The fusion rules \eqref{fusion} (see also Fig.~\ref{fusion-fig})
indicate that $h_{\lam_3}=h_{2\Lambda}$ ($k>1$) or $h_{\lam_3}=0$ 
($k\ge 1$). Then using the relation \eqref{weight}, one arrives at
$\Delta=1/(2(k+2))$ ($k>1$) or $\Delta=-3/(2(k+2))$ ($k\ge 1$). 
Correspondingly the partition functions are
\begin{equation}
Z_0=(x_1-x_2)^{-\frac{3}{2(k+2)}} \quad (k\ge1), 
\quad Z_2=(x_1-x_2)^{\frac{1}{2(k+2)}} \quad (k> 1).
\end{equation}
By inserting them into \eqref{sle-sum} and \eqref{driving-sum}, 
the driving processes $dx_{1t}$ and $dx_{2t}$ characterizing 
the geometric aspects of the SLE traces become:
\begin{align}
 dx_{1t}=\sqrt{a_1 \kappa }dB_{1 t}+\frac{2a_2+\kappa\Delta a_1}{x_{1t}-x_{2t}} dt, \quad
 dx_{2t}=\sqrt{a_2 \kappa }dB_{2 t}+\frac{2a_1+\kappa\Delta a_2}{x_{2t}-x_{1t}} dt,
\end{align}
where we normalized the variances by $dq_{\alpha t}=a_\alpha dt$ so that
$d \xi_{\alpha_t}=\sqrt{a_\alpha }dB_{\alpha t}$ with two independent
standard Brownian motions: $\mathbf{E}[d B_{\alpha_t}]=0$ and
$\mathbf{E}[dB_{\alpha_t} dB_{\beta t}]=\delta_{\alpha \beta} dt$.
These driving processes with the SLE \eqref{sle-sum} describe  two curves
emerging from two points $x_1=x_{1,0}$ and $x_{2,0}$.
Setting 
$y_{s}=x_{1t}-x_{2t}$ and rescaling the time  by $ds=\kappa(a_1+a_2)dt$,
we reduce the processes to the following Bessel process:
\begin{equation}
dy_s=d B_s+\frac{\Delta+2/\kappa}{y_s}ds
\end{equation}
with the effective dimension $d_{\rm eff}=2 \Delta+4/\kappa+1$. Substitution
of the exponent $\Delta$ and $\kappa$ \eqref{kappa} yields
\begin{equation}
d_{\rm eff}=\begin{cases}
            1 & \text{ ($k=1$)} \\
            \frac{2(k+1)}{k+2} &\text{ ($k>1$)} 
            \end{cases}
\quad \text{for $h_{\lam_3}=0$}, \quad
d_{\rm eff}=2+\frac{2}{k+2} \quad(k>1)
\quad \text{for $h_{\lam_3}=h_{2\Lambda}$}.
\end{equation}
Recalling that the Bessel process is recurrent (resp. not recurrent)
if $d_{\rm eff}<2$ (resp. $d_{\rm eff}>2$) (see \cite{Law}, for example), we conclude that
the driving processes $x_{\alpha t}$ hit each other with probability 1
for $h_{\lam_3}=0$ and never hit for $h_{\lam_3}=h_{2\Lambda}$. 
The hit of the driving processes means the hit of the tips of the SLE traces, and hence
the case for $h_{\lam_3}=0$ describes a single curve (i.e. single arch) 
whose end points are $x_1$ and $x_2$, while the case for $h_{\lam_3}
=h_{2 \Lambda}$ describes 
two curves converging to the point at $\infty$. The above 
observation agrees with Conjecture~\ref{conj-config}.

\subsection{Triple SLEs and arch (crossing) probabilities}
Let us discuss a more non-trivial example: the triple 
SLEs  ($m=3$). The geometric properties are characterized
by the partition function:
\begin{equation}
Z=\bra \psi_{\lam_4}(x_4)
\psi_{\lam_3}(x_3)\psi_{\lam_2}(x_2)\psi_{\lam_1}(x_1) \ket,
\end{equation}
where $x_4=\infty$ and by construction the bcc operators $\psi_{\lam_j}$ 
($1\le j\le m$) take values in  
the fundamental representation $L_\Lambda$. From the fusion 
rules (see also Fig.~\ref{fusion-fig}), the conformal weight 
$h_{\lam_4}$ of the bcc operator
$\psi_{\lam_4}(\infty)$ inserted at $\infty$ is 
$h_{\lam_4}=h_{3\Lambda}$ ($k\ge 3$) or $h_{\lam_4}=h_{\Lambda}$ 
($k\ge 1$). When the level takes its value in the range $k\ge 3$
and $h_{\lambda_4}=h_{3\Lambda}$,
there exists the factorized partition function with the same exponents
$1/(2(k+2))$:
\begin{equation}
Z=\left[(x_2-x_1)(x_3-x_1)(x_3-x_2)\right]^{\frac{1}{2(k+2)}} \quad (k \ge 3).
\end{equation}
According to  Conjecture~\ref{conj-config}, we expect
that this partition function describes  three curves
converging toward $\infty$. 
\begin{figure}[ttt]
\begin{center}
\includegraphics[width=0.7\textwidth]{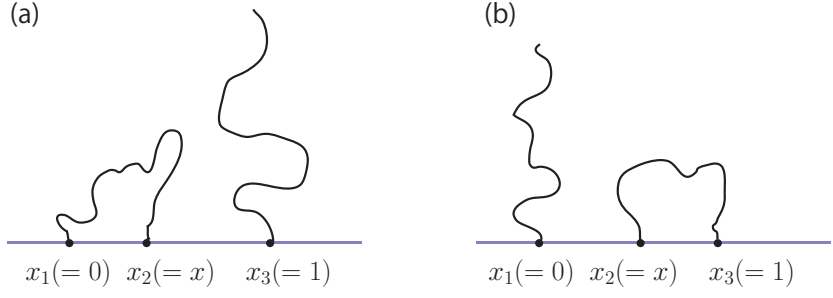}
\end{center}
\caption{Configurations for $m=3$ and $n=1$. 
There are two ($c_{3,1}=\binom{3}{1}-\binom{3}{0}=2$) topologically 
inequivalent configurations. These configurations are described by 
the partition function involving the 4 bcc operators with
the same conformal weights $h_\Lambda$. The configuration in the left (resp. right) 
panel is characterized by the partition function $Z_{\rm C_1}$ (resp. $Z_{\rm C_2}$). }
\label{config2}
\end{figure}

The case for $h_{\lam_4}=h_{\Lambda}$ is more interesting, because
two topologically inequivalent configurations do exist (see Fig.~\ref{config2}).
By conformal mapping $f(z)=(z-x_1)(x_3-x_4)/((z-x_4)(x_3-x_1))$,
the four points $x_j$ are transformed to the points $x_1\to0$,
$x_2\to x$, $x_3\to 1$ and $x_4\to\infty$, where $x=f(x_2)$. Thus the
partition function $Z$ can be expressed as
\begin{equation}
Z=((x_2-x_4)(x_1-x_3))^{-2 h_\Lambda}Z(x), \quad
Z(x)=\bra \psi_\Lambda(\infty) \psi_\Lambda(1) \psi_\Lambda(x) \psi_\Lambda(0) \ket.
\label{Z}
\end{equation}
The correlation function $Z(x)$ can be calculated by solving the KZ-equation 
\eqref{global2} \cite{KZ,FMS}. Its explicit form reads
\begin{equation}
Z(x)=Z_{\rm C_1}(x)+Z_{\rm C_2}(x)
\end{equation}
with
\begin{align}
Z_{{\rm C_1}}(x)=F_1^{(-)}+\frac{1-c_-}{c_+} F_1^{(+)}, \quad
Z_{{\rm C_2}}(x)=F_2^{(-)}+\frac{1-c_-}{c_+} F_2^{(+)},
\label{part}
\end{align}
where $F_j^{(\pm)}$'s are expressed by the hypergeometric function ${}_2F_1$: 
\begin{align}
&F_1^{(-)}=x^{-2 h_\Lambda}(1-x)^{h_{2\Lambda}-2h_{\Lambda}}
{}_2F_1\left(\frac{1}{k+2},\frac{-1}{k+2};\frac{k}{k+2};x\right), \nn \\
&F_1^{(+)}=x^{h_{2\Lambda}-2h_\Lambda}(1-x)^{h_{2\Lambda}-2h_{\Lambda}}
{}_2F_1\left(\frac{1}{k+2},\frac{3}{k+2};\frac{k+4}{k+2};x\right), \nn \\
&F_2^{(-)}=\frac{1}{k} x^{1-2 h_\Lambda}(1-x)^{h_{2\Lambda}-2h_{\Lambda}}
{}_2F_1\left(\frac{k+3}{k+2},\frac{k+1}{k+2};2\frac{k+1}{k+2};x\right), \nn \\
&F_2^{(+)}=-2 x^{h_{2\Lambda}-2h_\Lambda}(1-x)^{h_{2\Lambda}-2h_{\Lambda}}
{}_2F_1\left(\frac{1}{k+2},\frac{3}{k+2};\frac{2}{k+2};x\right), 
\end{align}
and the coefficients $c_\pm$ are respectively, given by
\begin{equation}
c_-=2\frac{\Gamma(2/(k+2))\Gamma(-2/(k+2))}{\Gamma(1/(k+2))\Gamma(-1/(k+2))}, \quad
c_+=-2 \frac{\Gamma^2(2/(k+2))}{\Gamma(3/(k+2))\Gamma(1/(k+2))}.
\end{equation}
Note that $Z_{\rm C_1}(x)$ and $Z_{\rm C_2}(x)$ satisfy the
relation  $Z_{\rm C_1}(x)=Z_{\rm C_2}(1-x)$.
For $k=1$, $c_-=1$ holds\footnote{Thus the second terms in \eqref{part} vanish. 
In fact, this is a 
consequence that primary fields with weight $h_{2\Lambda}$ do not exist for $k=1$.},
 and then the partition functions 
reduce to simple forms
\begin{equation}
Z_{{\rm C_1}}(x)=x^{-1/2}(1-x)^{1/2}, \quad Z_{{\rm C_2}}(x)=x^{1/2}(1-x)^{-1/2}.
\end{equation}
For generic $k$, the behaviors of the partition functions close to 
the position $x=0$ and $x=1$ are described as
\begin{equation}
Z_{{\rm C}_1}(x)\sim \begin{cases}
         x^{-2h_\Lambda} & \text{ $x\to0$} \\
         (1-x)^{h_{2\Lambda}-2h_\Lambda} &\text{ $x\to 1$}
         \end{cases},
\quad
Z_{{\rm C}_2}(x)\sim \begin{cases}
         x^{h_{2\Lambda}-2h_\Lambda} & \text{ $x\to0$} \\
         (1-x)^{-2h_\Lambda} &\text{ $x\to 1$}
         \end{cases}.
\end{equation}
Thus the fusion rules \eqref{fusion} and Conjecture~\ref{conj-config}
imply that $Z_{\rm C_1}$ describes two curves connecting the points 
$[0x]$ and $[1\infty]$
(configuration ${\rm C_1}$; see the left panel in Fig~\ref{config2})
while $Z_{{\rm C_2}}$ corresponds to the configuration $[\infty 0]$ and $[x1]$ 
(configuration ${\rm C_2}$; see the right panel in Fig~\ref{config2}).

\begin{figure}[ttt]
\begin{center}
\includegraphics[width=0.85\textwidth]{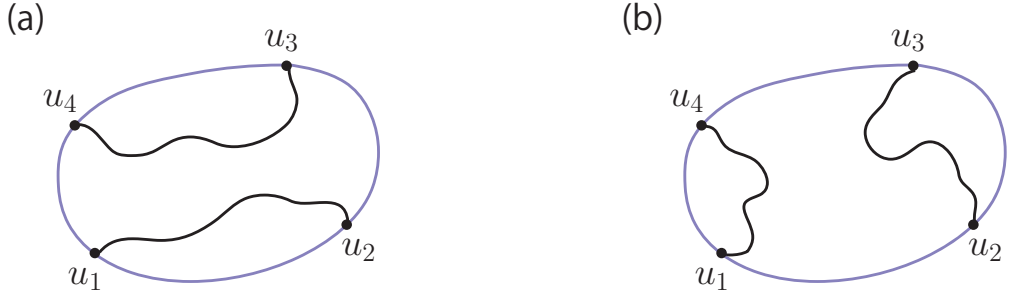}
\end{center}
\caption{Arch configurations in a domain $\mathcal{D}$ obtained by applying
a conformal transformations $u(z)$ to the configurations in Fig.~\ref{config2} 
($u_j$ denotes $u_j=u(x_j)$). 
The probability of 
the occurrence of the configuration (a) (resp. (b)) is  exactly the same as that 
of the occurrence of the configuration (a) (resp. (b)) in Fig.~\ref{config2}.}
\label{crossing-fig}
\end{figure}

\begin{figure}[ttt]
\begin{center}
\includegraphics[width=0.7\textwidth]{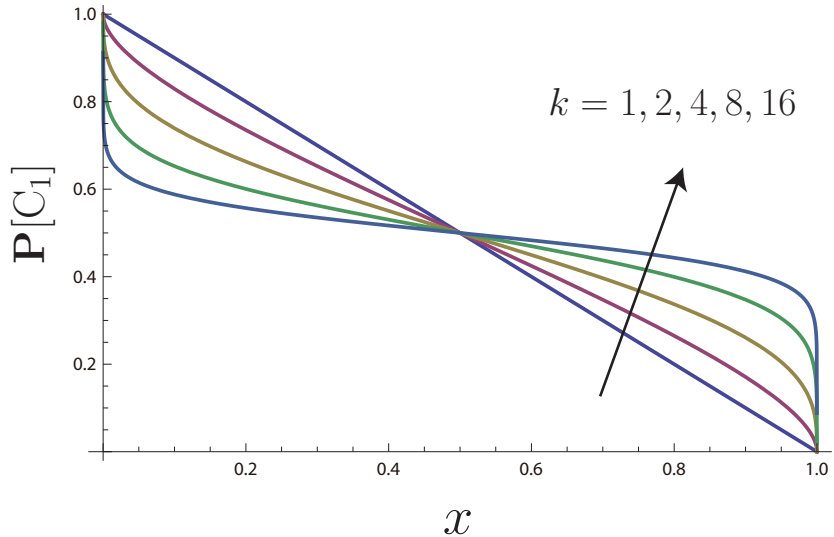}
\end{center}
\caption{The crossing probabilities $\mathbf{P}[{\rm C}_1]$
for various level $k$.}
\label{arch-fig}
\end{figure}

Utilizing the partition functions \eqref{part}, we can exactly compute the probability
$\mathbf{P}[{\rm C_1}]$ (resp. $\mathbf{P}[{\rm C_2}]$) of
the occurrence of the configuration ${\rm C_1}$ (resp. ${\rm C_2}$) with the
initial condition that the three curves emerging from the point $x_1=0$, $x_2=x$,
$x_3=1$ by
\begin{equation}
\mathbf{P}[{\rm C_1}]=\frac{Z_{\rm C_1}(x)}{Z_{\rm C_1}(x)+Z_{\rm C_2}(x)}, \quad
\mathbf{P}[{\rm C_2}]=\frac{Z_{\rm C_2}(x)}{Z_{\rm C_1}(x)+Z_{\rm C_2}(x)}.
\end{equation}
(See \cite{BBK,Koz} for similar formulas for minimal CFTs.)
Since the measure is conformally invariant, the probabilities in arbitrary (simply connected) 
domain $\mathcal{D}$ can be easily reproduced by use of  $Z(x)$ and the 
conformal map $u(z)$ mapping $\mathbb{H}$ to $\mathcal{D}$. More explicitly, the probability
$\mathbf{P}[{\rm C}_1]$ (resp. $\mathbf{P}[{\rm C}_2]$) is
exactly the same as that of the occurrence of the configuration with
two curves joining the boundary points $[u_1 u_2]$ and $[u_3 u_4]$ 
(resp. $[u_1 u_4]$ and $[u_2 u_3]$) on $\der \mathcal{D}$
(see Fig.~\ref{crossing-fig}) (note that $u_j$ stands for $u_j=u(x_j)$ $(1\le j \le 4$)).
 In this sense, the arch probabilities can be interpreted 
as crossing probabilities of the interfaces.
For $k=1$, the formula becomes very simple:
\begin{equation}
\mathbf{P}[{\rm C_1}]=1-x \quad (k=1).
\end{equation}
In Fig.~\ref{arch-fig}, the arch probabilities $\mathbf{P}[{\rm C_1}]$
are depicted for several values of $k$. As shown in Fig.~\ref{arch-fig},
the $x$-dependence of the arch probabilities close to $1/2$ in the 
wide range of $x$,  as increasing $k$. 

\section*{Acknowledgment}
The present work was partially supported
by Grants-in-Aid for Young Scientists (B) No. 21740285
and for Scientific Research (C) No. 24540393 from
Japan Society for the Promotion of Science.
%

\end{document}